\documentclass[10pt,conference]{IEEEtran} 

\usepackage{amsmath,amsthm,amsfonts,braket,graphicx,gensymb}

\begin{document}


\title{Finite Key Underwater Quantum Key Distribution: Performance Analysis and Improvements}

\author{\IEEEauthorblockN{Trevor Thomas}
\IEEEauthorblockA{\textit{School of Computing}\\
\textit{University of Connecticut}\\
Storrs CT, USA}
\and
\IEEEauthorblockN{Michael Warnock}
\IEEEauthorblockA{\textit{Naval Undersea Warfare Center}\\
Newport, RI USA\\
\textit{Department of Physics}\\
\textit{Brown University}\\
Providence, RI USA}
\and
\IEEEauthorblockN{Sarah Previte}
\IEEEauthorblockA{\textit{Department of Computer Science}\\
\textit{University of Alabama}\\
Tuscaloosa, AL USA}
\and
\IEEEauthorblockN{Benjamin Drozdenko}
\IEEEauthorblockA{\textit{Naval Undersea Warfare Center}\\
Newport, RI USA}
\and
\IEEEauthorblockN{Walter O. Krawec}
\IEEEauthorblockA{\textit{School of Computing}\\
\textit{University of Connecticut}\\
Storrs CT, USA\\
\texttt{walter.krawec@uconn.edu}}}

\maketitle

\begin{abstract}
	Quantum key distribution (QKD) allows for the establishment of a secret key between two parties, secure against computationally unbounded adversaries.  Most experimental and theoretical research in this area, investigates the performance of QKD over fiber or free space channels.  A growing body of work, however, has begun to investigate its performance in underwater scenarios.  Here, we consider the realistic finite key scenario, and evaluate the simulated performance of two different decoy state protocols.  We show that there are certain underwater channels where ``simpler'' QKD protocols outperform more complex ones.  Finally, we also investigate methods to improve QKD performance underwater, specifically looking at classical advantage distillation, which we show can greatly improve the maximal distance supported by QKD underwater.
\end{abstract}

\section{Introduction}
Quantum key distribution (QKD) is, by now, a well studied field, both in theory and in practice.  QKD protocols allow for the establishment of a secure secret key between two parties, referred to as Alice and Bob.  The security of such a system is based on physical assumptions (in particular, the postulates of quantum mechanics), as opposed to computational assumptions.  As it turns out, any key distribution protocol using classical communication, alone, can only be secure under computational assumptions --- that is, one must always rely on computationally hard problems to gain security using classical communication for key distribution tasks.  QKD does not share this limitation, though it does introduce its own set of challenges, particularly in terms of distance and noise tolerance limitations.  Such problems are exacerbated in an underwater scenario that we consider here, due to the high loss, and high noise, of underwater quantum channels.  For a general survey of quantum key distribution, the reader is referred to \cite{pirandola2020advances,amer2021introduction}.

While QKD is well studied when operating over fiber or even free-space channels (e.g., satellite communication), it has been discovered that QKD may be a very capable solution to underwater secure communication.  Several works have developed the theory behind underwater QKD and there have also been recent experimental implementations.  We will review prior work in greater detail in Section \ref{sec:pw}.  For a general survey of underwater QKD, the reader is referred  to \cite{paglierani2023primer}.

In this paper, we simulate, for the first time to our knowledge, the decoy state BB84 protocol \cite{hwang2003quantum,wang2005beating,lo2005decoy} in finite time, that is, in the practical finite key scenario.  This is opposed to the asymptotic scenario, investigated in previous theoretical work.  While asymptotic limits are important to understand the ``best case'' performance of QKD, it is vital to also simulate finite key performance to get an idea as to how these systems may perform in practice.  In particular, we show that in the finite key setting, maximal tolerated distances are not nearly as high as that reported by the asymptotic evaluations.  We also show that, in the finite key scenario, systems struggle to produce a secret key when there is a non-trivial amount of ambient light at sea-level (making it potentially practical at night, but in day-light conditions, QKD may struggle).  However, we also investigate ways to improve performance under these conditions, as we discuss shortly.

As a second contribution, we investigate an alternative ``simpler'' decoy-state protocol, which requires fewer decoy settings than the standard ``full'' decoy state protocol.  To our knowledge, this protocol has not been investigated in underwater settings (either in the finite key or in the asymptotic scenario).  We show that this protocol can outperform the standard decoy state protocol under some operating conditions, while in others it performs worse.  Understanding the conditions for when one protocol should be used over another is vital for future practitioners of this technology.  Furthermore, one may ``switch'' from one protocol to another based on factors such as depth, distance, ambient light (e.g., time of day), and water quality.  Our work shows that protocol choice is vital for optimizing performance in terms of overall secret key rates, and the effects underwater can be substantial.

As a final contribution, we also investigate the performance of Classical Advantage Distillation (CAD) techniques \cite{maurer2002secret,bae2007key}.  These methods have been shown to benefit fiber-based QKD implementations, but have not been investigated in underwater simulations.  We show that CAD can greatly benefit key-rate performance, in several scenarios, allowing for continued operation of QKD underwater, for longer distances.  In particular,  CAD can produce a secret key underwater when normal operation would fail.  It is especially useful, when there is a higher level of ambient light at sea level.  We investigate both the standard and the ``simple'' decoy state protocol with CAD, showing when CAD can help, and when it can degrade performance.

Ultimately, our work shows that QKD may operate underwater in the practical finite key scenario, though as expected, maximal distances are not as optimistic as asymptotic simulations have shown in prior work.  We use standard underwater communication models (discussed in detail in Section \ref{sec:uw}), allowing us to simulate the behavior of a QKD system operating in a variety of water conditions.  We discover that a simpler decoy state protocol may actually outperform the ``standard'' version in certain operating conditions.  Finally, we demonstrate the importance of investigating alternative post-processing methods for improving overall key generate rates.  In particular, we show that CAD can greatly improve the performance of QKD underwater, and we evaluate its behavior in a variety of operating scenarios.

\subsection{Prior Work}\label{sec:pw}

We distinguish, here, between theoretical studies, which involve analyzing underwater QKD performance through simulations using mathematical models of underwater quantum communication (generally modeling attenuation and background noise through a simulated photonic implementation of a BB84 system), and between experimental studies which actually implement QKD in an underwater environment, either in a lab setting, or in a natural body of water.  Our work in this paper is purely on the theoretical side, using mathematical models derived through various sources discussed below.  However, we also discuss several experimental results to put our own results, shown later, into better context.  We note that this section is not meant to be a complete survey; instead, we focus primarily on recent work most related to our own.  For a general survey of underwater QKD developments, the reader is referred to a recent survey paper \cite{paglierani2023primer}.

Perhaps the first author to investigate theoretical underwater QKD was Lanzagorta in his 2012 book \cite{lanzagorta2012underwater}.  Since then, several other works have investigated the theoretical performance of underwater QKD.  For instance, \cite{lanzagorta2019assessing}, studied the performance of QKD between a surface vessel and a submerged receiver, using mathematical models of the noise derived in \cite{rogers2006free}.  Their result claimed that QKD may operate at depths up to $60m$, in Jerlov Type I ocean water, though this distance dropped substantially as the water clarity decreased (for more information on the Jerlov water scale, originally proposed in 1951, see \cite{jerlov1976marine,solonenko2015inherent}).  Analyses using Monte Carlo simulations were also performed in \cite{shi2015channel} and show similar distance results.  Numerical analyses of QKD over the air-water interface were also performed in \cite{zhou2015performance}.

Theoretical studies involving horizontal quantum communication between two submerged objects were performed in \cite{zhao2019performance,lopes2018optimized,gariano2019theoretical} where the decoy state BB84 \cite{hwang2003quantum,wang2005beating,lo2005decoy} protocol was used (we discuss this protocol, later, in Section \ref{sec:protocol}).  The decoy state protocol is the protocol we will be analyzing; we will also be considering the case of ``horizontal'' communication between two submerged objects.  These theoretical studies suggest distances of up to $100m$ are possible depending on the water clarity.  We note that these studies, and the ones discussed above, only considered the asymptotic theoretical upper-bounds on QKD performance, not the realistic finite-key performance of a QKD system, the latter of which generally performs worse, yet results in a more accurate model, from a practical standpoint.  In this work, we will consider the finite-key performance of decoy-state BB84  in a simulated underwater channel for the first time.

On the experimental side, perhaps the first experimental study of the feasibility of underwater QKD was performed in 2017, in \cite{ji2017towards}.  There, an experiment was performed involving a tube, $3.3m$ in length, filled with Jerlov Type I seawater.  In this experiment, a relatively high fidelity of $98\%$ was observed.  High dimensional implementations, using twisted photons, were performed in an outdoor pool in \cite{bouchard2018quantum}.  The pool was placed outside and had temperature variations between $17\degree C$ and $27\degree C$.  This experiment showed QKD was feasible under these lab conditions for a distance of $3m$.  An experiment was performed in \cite{hufnagel2019characterization} involving quantum communication through the Ottawa River in Canada for a distance of $5.5m$.  Finally, in \cite{hufnagel2020investigation}, a distance of $30.5m$ was achieved in an experiment involving clear water.

\subsection{QKD Basics}

Users of QKD protocols require access to two communication channels: a quantum channel, capable of sending quantum states (ideally qubits) from one party to another, and a classical authenticated channel, capable of sending classical messages in an authenticated manner (but not secret).  The authentication is typically implemented using an information theoretic secure Message Authentication Code (MAC).  The quantum channel has no authentication or secrecy (beyond that which is provided by the laws of quantum mechanics).  The classical channel may be read by an adversary, but not written to by an adversary.  As is typical in QKD literature, we will refer to the two honest, communicating, parties as Alice and Bob (the parties who wish to establish a shared secret key); we will refer to the adversary as Eve.

QKD protocols operate in two stages: first, Alice and Bob utilize the quantum and classical channels to establish a classical \emph{raw key} during the so-called \emph{quantum communication stage} of the QKD protocol.  This raw key is partially secret (an adversary may have some information on it based on her attack) and it will be noisy (natural and adversarial noise will create disturbances in Alice and Bob's respective raw keys).  Thus, this raw key cannot be used immediately as a secret key and must be further processed.

The second stage is purely classical and consists of a pre-processing stage (which is optional) to produce new raw keys (or the same if pre-processing is not used).  Next, error correction and privacy amplification are run.  Error correction leaks additional information to the adversary, however it results in a fully correlated raw key (with high probability); privacy amplification takes this $n$-bit error corrected raw key, and outputs a smaller $\ell$-bit secret key.  This is done by choosing a random two-universal hash function and applying it to the raw keys.  See \cite{renner2008security} for more details.

One important metric of QKD performance is the \emph{key rate} of the system, defined to be the ratio of $\ell$ (the secret key size) to the total number of signals sent during the quantum communication stage.  In general, this is a function of several factors, including the noise and loss rate of the channel, along with the performance of any additional, optional, classical post-processing.  In the finite key setting, it also depends on how many rounds are used for sampling, or testing the fidelity of the channel. In general, the larger the loss rate and the higher the noise, the smaller $\ell$ will be, thus the lower the efficiency of the system will be.  The larger $N$ is, the more rounds that can be used for sampling, thus increasing the confidence of the noise bounds, and leading to a larger final key size, $\ell$.


The physical channel characteristics (e.g., noise and loss), however, are not the only concern in a practical QKD implementation.  Practical device imperfections can also affect the performance of the system.  For instance, detectors may have false positive or negatives, leading to decreases in key-rate.  Furthermore, and even more problematic, Alice's source device is incapable of sending single qubits.  In general, quantum states are encoded as photons (we will assume polarization encoding later).  For this, Alice uses a weak coherent source to produce a signal.  Such a source emits $m$ photons with Poisson distribution $p(m|\mu)$, where $\mu$ is the ``power level'' of the source (the mean photon number).  Whenever Alice attempts to emit a photon, she may set (and alter) $\mu$, but she never actually knows how many photons $m$ are leaving her lab in any particular instance.

This is problematic for QKD as, in general, whenever Alice emits two or more photons, Eve, the adversary, may keep one of them and later learn everything about this particular round.  Whenever Alice emits zero photons, Eve learns nothing, but of course Bob also learns nothing (if he detects anything, it is random noise).  A ``good round'' can only occur when $m=1$.   In general, the key-rate will depend on the single-photon gain (the probability that Alice and Bob distill a raw key bit in the event Alice sent a single photon) and the single-photon error rate \cite{gllp}.  Both of these are not directly observable.  However, the decoy state method, discussed in Section \ref{sec:protocol}, can be used to derive tight bounds on these necessary quantities.

\section{Underwater Quantum Communication}\label{sec:uw}

We now describe the basic model for underwater qubit transmission, including noise and loss, that we will use in our work.  This model is from previous work and we only summarize it here; for more details, the reader is referred to \cite{paglierani2023primer} and references therein.

Consider a weak coherent source used at Alice's end.  This device emits photons with a Poisson distribution.  Namely, a power level $\mu$ is specified by the user, and the source will emit $m$ photons ($m\ge 0$), with probability:
\begin{equation}
	p(m|\mu) = \frac{e^{-\mu}\mu^m}{m!}
\end{equation}
These photons are then polarized at an angle specified by Alice (chosen randomly according to the protocol under consideration), and travel towards a  receiver which has a measurement apparatus consisting of an optical switch (to choose one measurement basis or another), and several single photon detectors (SPDs) to observe incoming photons.  These SPDs are not able to count photons; instead they ``click'' when one or more photons hits the device.  Practical SPDs suffer from dark-count rates (a click when no photon hits the detector) and also non-unit efficiency (i.e., they fail to click even if a photon hits it).  Both must be taken into account when modeling the channel characteristics.  

In addition to the source device, there is also background light hitting the detector.  These background photons introduce noise into the signal as they will randomly cause detectors on Bob's side to click.  The number of background photons reaching the detector depend, primarily, on the background light level at the water's surface, the depth of the receiver, and the angle of the receivers aperture.

In this work, we will assume the source and receiver are at the same depth, thus photons must travel horizontally between parties.  The two parties are located apart from one another $L$ meters in distance.  There are two critical elements we must model: First is the \emph{transmittance} of the channel, $\gamma$ (where, if Alice sends $N$ photons to Bob, $\gamma N$ will arrive on average).  Second is the total number of background photons arriving at Bob's device (which will introduce noise).  We assume background photons will cause random detector errors in the receiver's end.  Once both of these characteristics are defined, we later develop a simulator which will simulate the effects of the channel on a basic QKD implementation.  

\subsection{Attenuation Loss}
We first discuss the model for loss in the channel.  Let $d_{TX}$ be the diameter of the exit pupil of the source and $\lambda$, the wavelength of the transmitted photons.  As discussed, $L$ is the distance (in meters) between the source and receiver (which are at the same depth).  See also Table \ref{tbl:notation} for a table of the important notation used in the model.  The discussion in this section is derived from \cite{paglierani2023primer} and references therein; we only include a brief summary, here, for completeness.

Attenuation loss is affected by several factors, including scattering, turbulence, and absorption.  To model $\gamma$, after the source emits a signal with power $P_T$ from its pupil $P_0$ (of diameter $d_{TX}$), the pattern sent can be described by $\sqrt{P_T}\mu_0(r)$, for $r\in P_0$ such that $\int_{P_0}|\mu_0(r)|^2dr = 1$.  At the receiver's end, the received pattern, for some $\rho \in P_1$ (where $P_1$ is the receiving pupil of diameter $d_{RX}$), is denoted $\mu(\rho)$, and can be found to be:
\begin{equation}
	\mu(\rho) = \sqrt{P_T}\int_{P_0}\mu_0(r)h(r,\rho)dr.
\end{equation}
Above, $h(r,\rho)$ is the underwater quantum channel impulse response for a waveform propagating through some turbulent channel \cite{shapiro2003near}.  This is defined to be:
\begin{align}\label{eq:1}
	h(r, \rho) &= \sqrt{A(L)}\times\frac{\exp\left(ikL+ \frac{ik(r-\rho)^2}{2L}\right)}{i\lambda L}\notag\\
&\times \exp\left(\psi(r,\rho) + i\chi(r,\rho)\right).
\end{align}
Above, $A(L)$ is defined to be:
\begin{equation}
	A(L) = \exp\left(-\alpha L \left[ \frac{d_{RX}}{\theta L}\right]^T\right),
\end{equation}
where $\alpha$ is the attenuation coefficient which depends on the water type (discussed shortly); $\theta$ is the beam divergence angle (the angle for which the beam intensity drops to $1/e$ of its maximal value); and $T$ is a correction parameter depending on the water type \cite{elamassie2018performance}.

For the attenuation coefficient,  $\alpha$, there are two primary ways to determine a reasonable setting outside of setting up an actual experimental implementation.  First is the Jerlov classification system \cite{jerlov1976marine,solonenko2015inherent}, and the second is the Mobley classification \cite{mobley1994light}.  The first splits water into two main groups: \emph{Open Ocean} and \emph{Coastal}.  Open Ocean is further divided into types I, IA, IB, II, and III, while Coastal is divided into types $1, 3, 5, 7$ and $9$.  In this scale, the overall clarity of the water decreases as the scale number increases.  The Mobley classification has a higher range of coefficients, but is only applicable to the blue/green wavelength (which is suitable for QKD as that is the wavelength used).  In our work, we will consider  certain coefficient values based on the Mobley scale, shown in Table \ref{tbl:water}.

\begin{table}
\caption{Attenuation coefficients ``$\alpha$'' for a select number of water types, according to the Mobley scale.  These values were taken from \cite{paglierani2023primer}.}\label{tbl:water}
\centering
\begin{tabular}{lc}
\hline
	Water type & Setting for $\alpha$\\
\hline
	Pure sea water & 0.043\\
	Clear ocean water & 0.151\\
	Coastal ocean water & 0.398\\
	Turbid harbor water & 2.190\\
\hline
\end{tabular}
\end{table}

Returning to Equation \ref{eq:1}, the function $h(r,\rho)$ can be written in its functional SVD form as:
\begin{equation}
	h(r,\rho) = \sqrt{A(L)}\sum_{j \ge 1} \sqrt{\mu_j}f_j(r)\phi_j(\rho),
\end{equation}
where $1 \ge \mu_1 \ge \mu_2 \ge \cdots \ge 0$ are the \emph{modal transmitivities}.  In general, $\mu$, $f_j$, and $\phi_j$ are random variables; ``however it was shown in \cite{shapiro2003near,fahim2020performance} that one can lower bound the expected value of $\mu_1$ (the largest) by the following $\mu_{turb}$:
\begin{align}
	\mu_{turb}&= \frac{8\sqrt{F}}{\pi}\int_{0}^1 \exp\left(-\frac{1}{2}W\left(d_{TX} x, L\right)\right)\notag\\
	&\times \left(\cos^{-1}(x) = x\sqrt{1-x^2}\right) J_1(4x\sqrt{F})dx,
\end{align}
where $F$ is the Fresnel number:
\begin{equation}
	F = \left(\frac{\pi d_{TX}d_{RX}}{4\lambda L}\right)^2,
\end{equation}
and where $J_1$ is the first-order Bessel function of the first kind.  The function $W(\rho, L)$ is the wave structure function defined in \cite{fahim2020performance,raouf2022performance} to be:
\begin{align}
	W(\rho, L) &= 1.44\pi \left(\frac{2\pi}{\lambda}\right)^2L \epsilon^{1/3}\left(\frac{\alpha_{th}^2\chi_T}{\omega^2}\right)\notag\\
	&\times \left(1.175\eta_K^{2/3}\rho + 0.419\rho^{5/3}\right)\notag\\
	&\times \left(\omega^2 + d_r - \omega(d_r+1)\right).
\end{align}
The above introduced several new variables which are defined in Table \ref{tbl:notation}. Default values (for simulation purposes) for all parameters can also be found in that table.
Above, the eddy diffusivity ratio, $d_r$, can be approximated by   \cite{elamassie2017effect}:
\begin{equation}\label{eq:dr}
d_r = \frac{|\omega|}{R_F}
\end{equation}
 where:
\begin{equation}
	R_F = \left\{\begin{array}{ll}
		|\omega| - \sqrt{|\omega|(|\omega| - 1)} & \text{ if } |\omega| \ge 1\\
		1 / (1.85 - 0.85|\omega|^{-1}) & \text{ if } 0.5 \le |\omega| \le 1\\
		1/0.15 & \text{ otherwise}
		\end{array}\right.
\end{equation} 
The value of $\omega$, which is the relative strength of the temperature and salinity fluctuations, is itself a function of the thermal expansion coefficient $\alpha_{th}$, and also the saline contraction coefficient \cite{elamassie2017effect}, both of which can be computed using the TEOS-10 software\footnote{\texttt{teos-10.org/software.htm}}.  We will use standard default values used in other theoretical underwater studies, shown in Table \ref{tbl:notation}, in order to better compare with prior work.

\subsection{Noise}
To simulate noise in the channel, we must determine the average number of background photons arriving at the receiver, along with the effect of dark counts of the receiver's detectors.  As with loss, discussed above, the information in this section is taken from the survey \cite{paglierani2023primer}.

The number of noise photons (excluding dark counts which we will incorporate into our simulator directly, as we discuss in Section \ref{sec:sim}) is computed to be:
\begin{equation}\label{eq:num-noise}
	n_N = \frac{1}{2}\frac{\pi E_d A \Delta t'\lambda\Delta\lambda(1-\cos(\Omega))}{2h_pc}.
\end{equation}
Above, $A$ is the receiver aperture area (itself, based on $d_{RX}$, namely $A = \frac{\pi}{4}d_{RX}^2$); $\Omega$ is the field of view of the detector; $\Delta t'$ is the receiver gate time; $\Delta t$ is the bit period; $\lambda$ is the wavelength; $h_p$ is Planck's constant, and $c$ is the speed of light.  Finally, $E_d$ is the irradiance of the environment, given by $E_d = E_0\exp(-K_\infty \zeta)$, where $E_0$ is the irradiance at the sea surface; $K_\infty$ is the asymptotic value of the spectral diffuse attenuation coefficient \cite{mobley1994light}; and finally $\zeta$ is the underwater depth.  Again, see Table \ref{tbl:notation} for a list of default values that we will use later for these many  model parameters.

We will assume that all noise photons are randomly polarized; thus half (in expectation) will travel towards the receiver's ``zero'' detector while the other half will travel towards the ``one'' detector.  We will discuss this further in Section \ref{sec:sim} when we discuss our simulation and evaluation techniques.  

\begin{table*}[t]
\caption{Table of Notation used by the underwater model, along with default values used later in our simulations (these defaults are standard values for the majority of simulated underwater quantum communication \cite{paglierani2023primer}).}\label{tbl:notation}
\centering
\begin{tabular}{lll}
\hline
	Parameter & Description & Default value\\
	\hline
	$d_{TX}$ & Diameter of source exit pupil & $30cm$\\
	$d_{RX}$ & Diameter of receiver pupil & $30cm$\\
	$\lambda$ & Wavelength of transmitted photons & $530nm$\\
	$L$ & Distance (meters) between source and receiver & Variable\\
	$\alpha$ & Attenuation Coefficient based on Jerlov or Mobley scale & Variable\\
	$T$ & Correction parameter & 0.26\\
	$\theta$ & Beam divergence angle & $6$ degrees\\
	$\alpha_{ch}$ & Thermal expansion coefficient & $2.56\times 10^{-4}$\\
	$\chi_T$ & Dissipation rate of mean-squared temperature & $10^{-5}$\\
	$\omega$ & Relative strength of the temperature and salinity fluctuations & $-2.2$\\
	$\epsilon$ & The dissipation rate of turbulent kinetic energy&$10^-5$\\
	$\nu$ & Kinematic viscosity&$1.0838\times 10^{-4}$\\
	$\eta_K$ & Kolmogorov microscale length &$\eta_K= (\nu^3/\epsilon)^{1/4}$\\
	$d_r$ & Eddy diffusivity ratio & See Equation \ref{eq:dr}\\
	$\Delta t$ & Bit period & $35ns$\\
	$\Delta t'$ & Receiver gate time & $200ps$\\
	$\Omega$ & Field of view & $180$ degrees\\
$E_0$ & irradiance of the environment at the sea surface& Variable\\
\hline
\end{tabular}
\end{table*}
\section{The Protocol}\label{sec:protocol}

As discussed earlier, the important metric in QKD evaluation is its secret key rate.  Past theoretical underwater QKD work, to our knowledge, only evaluated asymptotic upper bounds.  To evaluate more practical finite key results, we will use key-rate derivations from \cite{lim2014concise} (for the standard three-decoy protocol) and \cite{rusca2018finite} (for the simpler two-decoy protocol); both of these, to our knowledge, represents the state of the art for finite key decoy state BB84 over arbitrary channels.  This derivation abstracts away the noise and loss as parameters.  We will then use a simulator (discussed, below, in Section \ref{sec:sim}) which will model the appropriate loss and noise parameters as discussed above, needed to evaluate the key rate.  

The decoy-state BB84 protocol introduced in \cite{hwang2003quantum,wang2005beating,lo2005decoy} requires Alice to alternate the power level of her photon source, in order to obtain better statistics on the channel.  Specifically, to compute the secret key rate of a QKD protocol, we require knowledge of the single photon yield and error rate.  Normally, one is forced to use pessimistic bounds on these values, as they are not directly observable.  However, the decoy state protocol is able to derive very tight bounds on these parameters, by altering, randomly, the source power level between rounds.

We assume Alice has a weak coherent source, capable of sending $m$ photons with probability $p(m|\mu)$, with $\mu$ being set by Alice.  Though Eve can count the number of photons on any particular round (and thus adapt her attack based on the photon number), the setting of $\mu$ for any particular round is kept secret, and Eve cannot distinguish between different settings of $\mu$ (which Alice will choose randomly every round).  

The decoy state protocol operates as follows:

\textbf{Public Parameters:}
\begin{enumerate}
\item Intensity Settings $\mathcal{P} = \{\mu_1, \mu_2, \cdots\}$.
\item Intensity Setting Distribution: $\{p_1, p_2, \cdots\}$, where $p_i$ is the probability Alice chooses $\mu_i$ on any particular round.
\item Basis Bias Choice: $p_Z$ (the probability that parties  choose the $Z$ basis).
\item Number of rounds: $N$.
\end{enumerate}

\textbf{Quantum Communication Stage:}
\begin{enumerate}
  \item Alice prepares, using a weak coherent source, a random $Z$ basis state, with probability $p_Z$; otherwise, she prepares a random $X$ basis state.  This is done using  a random intensity choice $\mu_i$ with probability $p_i$.
\item Bob chooses to measure in the $Z$ basis with probability $p_Z$, otherwise the $X$ basis.  This leads to one of four possible outcomes: $0, 1, v, $ or $d$
  \begin{itemize}
  \item $0$ (meaning he observed a zero state, either $\ket{0}$ or $\ket{+}$)
  \item $1$ (meaning he observed a one state, either $\ket{1}$ or $\ket{-}$)
  \item ``$v$'', meaning he observed a vacuum (i.e., nothing)
\item  ``$d$'' meaning he observed a double click (both his zero and one detectors clicked).
\end{itemize}
\item The above repeats $N$ times, after which Alice discloses her intensity choices, while both Alice and Bob disclose their basis choice.  This allows Alice and Bob to determine integers $N_{X, i}$ and $N_{Z,i}$ which are the number of non-vacuum events where both parties used the $X$ (respectively $Z$) basis and where Alice chose intensity setting $\mu_i$.
\item For all $Z$ basis rounds, Alice discloses her state choice and Bob discloses his measurement outcome.  This allows Alice and Bob to determine $M_{Z,i}$, which is the number of times Bob observed a value different from what Alice sent when she initially used power setting $\mu_i$.  Note that $Z$ basis rounds are used for channel tomography, while $X$ basis rounds will be used for the key.
\item Parties output a raw key bit for every round where parties chose the $X$ basis and where Bob did not see a vacuum: Alice will use her initial state choice while Bob will use his measurement outcome.  In the event a particular round was a double event ($d$), he will choose zero or one randomly for his key bit.
\end{enumerate}

\textbf{Classical Post-processing Stage:}
\begin{enumerate}
\item Parties take their raw keys and optionally perform a Classical Advantage Distillation process (discussed in Section \ref{sec:cad}).  This is optional.
\item Parties next run an error correction protocol which leaks $\lambda_{EC}$ bits of information.
\item Finally, parties run Privacy Amplification which hashes the error corrected raw key to a secret key of size $\ell$ bits. 
\end{enumerate}

An important question is to determine the size of the final secret key $\ell$.   This will depend on the channel parameters as well as the parameters Alice and Bob may set.  We will first consider the case without the optional CAD process; key-rates for when CAD is activated will be discussed in Section \ref{sec:cad}.

We consider two different decoy protocols in this work: the ``standard'' three-decoy state protocol where three possible power level settings are used, one of which is set close to zero so that Alice emits a vacuum with high probability; and the ``two decoy'' state protocol, where only two power level settings are used (neither of which are set to near vacuum).

We will show later that sometimes one protocol outperforms the other in underwater channels.  We comment that this is the first time the two-decoy state protocol has been investigated for underwater quantum communication (either asymptotically or in the finite key setting).  Note that the two-decoy protocol is sometimes called the one-decoy protocol since, traditionally, a distinguished power level was designated as the ``signal'' power level, while any additional settings were ``decoys.''  However, new security proofs utilize all power levels; thus, we use $n$-decoy to mean there are $n$ choices of power level settings to avoid confusion here.

\subsection{Three-Decoy State Key-Rate Bound}
For the standard three-decoy state protocol, we may use results in \cite{lim2014concise} to determine the size of $\ell$ as:
\begin{equation}\label{eq:key-size}
\ell = n_0 + n_1(1 - h(\phi_X)) - \lambda_{EC} - 6\log_2\frac{21}{\epsilon} - \log_2\frac{2}{\epsilon},
\end{equation}
where $\epsilon$ is a user-specified security parameter, $\lambda_{EC}$ is the error correction leakage, and $n_i$ is the estimated number of rounds which did not lead to a vacuum event and where Alice initially sent $i$ photons.  These can be bounded by:
\begin{equation}
n_0 = \tau_0\frac{\mu_2 N_X(\mu_3, 1) - \mu_3 N_X(\mu_2, 0)}{\mu_2 - \mu_3}
\end{equation}
and:
\begin{equation}\label{eq:n1}
n_1 = \tau_1\mu_1\frac{N_X(\mu_2, 1) - N_X(\mu_3, 0) - \frac{\mu_2^1 - \mu_3^2}{\mu_1^2}(N_X(\mu_1,0) - \frac{n_0}{\tau_0})}{\mu_1(\mu_2-\mu_3) - \mu_2^1 + \mu_3^2}.
\end{equation}
Above, we define $\tau_x$ to be the probability that Alice sends $x$ photons, averaged over all intensity settings, namely:
\begin{equation}
\tau_x = \sum_{j=1}^3 \frac{e^{-\mu_j}\mu_j^xp_j}{x!},
\end{equation}
and where:
\begin{equation}\label{eq:nxp}
N_X(\mu, i) = \frac{e^{\mu}}{p_\mu}\left(N_{X,\mu} + (-1)^i \sqrt{\frac{N_X}{2}\ln\frac{21}{\epsilon}}\right).
\end{equation}
Above, $N_X = \sum_j N_{X,j}$, where $N_{X,j}$ is determined by the protocol execution.  Note that we abuse notation, slightly, above and write $N_{X,\mu}$ to mean $N_{X,j}$, where $j$ is the index where $\mu_j = j$.

This bounds the single and vacuum yields; what remains is a bound on the \emph{phase error} term, namely $\phi_X$.  This will utilize statistics $m_Z$ which, recall, is the number of error events in the $Z$ basis (an observable quantity since $Z$ basis rounds are used for error checking and will be completely divulged).  Then:
\begin{equation}
\phi_X = \frac{v}{z_1} + \gamma\left(\epsilon, \frac{c}{z_1}, z_1, n_1\right).
\end{equation}
where $z_1$ is defined similarly to Equation \ref{eq:n1}, but using $Z$ basis statistics, instead of $X$ basis, and:
\begin{equation}
v = \tau_1\frac{m_{Z}(\mu_2,0) - m_{Z}(\mu_3,1)}{\mu_2-\mu_3}.
\end{equation}
Above, the function $m_Z(\mu,i)$ is defined similarly to Equation \ref{eq:nxp} but, again, using the $Z$ basis statistics.  Finally, $\gamma$ is the following term, based on sampling error:
\begin{equation}
\gamma(a,b,c,d) = \sqrt{ \frac{(c+d)(1-b)b}{cd\ln 2} \log_2\left( \frac{c+d}{cd(1-b)b}\frac{21^2}{a^2}\right)}
\end{equation}

See \cite{lim2014concise} for further details on the above derivations.

\subsection{Two-Decoy State Key-Rate Bound}

For the two-decoy protocol (where Alice can only choose from two settings, neither of which is a vacuum setting), similar expressions may be found in \cite{rusca2018finite}.  In particular, the key-length is:
\begin{equation}\label{eq:key-size-2}
\ell = n_0 + n_1(1 - h(\phi_X)) - \lambda_{EC} - 6\log_2\frac{19}{\epsilon} - \log_2\frac{2}{\epsilon},
\end{equation}
Note the only difference in the structure of the above, compared to Equation \ref{eq:key-size} is the different constant of $19$ in the logarithm.  The main different is in how the yields and phase error terms are bounded.  Namely, now, we have:
\begin{equation}
n_0 = 2\left(\tau_0\frac{e^k}{p_k}\left(m_{Z,k} + \sqrt{\frac{m_Z}{2}\log\frac{1}{\epsilon}}\right) + \sqrt{\frac{n_Z}{2}\log\frac{1}{\epsilon}}\right)
\end{equation}
and:
\begin{equation}
n_1 = \frac{\tau_1\mu_1}{\mu_2(\mu_1-\mu_2)} \left( n_{Z,\mu_2}^- - \frac{\mu_2^2}{\mu_1^2}n_{Z,\mu_1}^+ - \frac{(\mu_1^2 - \mu_2^2) n_0}{\mu_1^2 \tau_0}\right)
\end{equation}
Since $n_0$ depends on the power setting $k$, we minimize the final key length (Equation \ref{eq:key-size-2}) over both power levels.   The phase error bound is computed as described in the previous section.

\section{Simulation}\label{sec:sim}

We evaluate the performance of the above QKD protocols, in the finite key setting, by simulating, round by round, the underwater channel.  This will allow us to observe the performance of the overall QKD system as time (in our case rounds of the protocol) increases.

In particular, for every round of the QKD protocol (where the total number of rounds, $N$, is set by the user), we perform the following processes in sequence:
$ $\newline
\textbf{Initial Randomness and Source Preparation:}
\begin{enumerate}
  \item First, choose a random basis choice for Alice and Bob.  These choices are independent of one another, and basis $Z$ is chosen with probability $p_Z$ set by the user. 
\begin{itemize}
\item If Alice and Bob choose different bases, we skip this round, and repeat this step.  We do, however, count the total number of rounds, \emph{including} those which were skipped for the final key-rate computation.  This is due to the fact that mismatched bases lead to ``lost time'' in the sense that no key is distilled, however users must, in practice, go through the entire process before realizing they chose the wrong basis.  For the sake of simulation, however, there is no point in continuing in this event, other than to keep track of these rounds in the running total.
\end{itemize}
  \item Next, choose a random power level for the source (Alice) to use for this round.  In this case, one of the $k$ power levels, $\mu_k$ will be chosen with the corresponding probability $p_k$ as discussed in Section \ref{sec:protocol}.
  \item From the above, choose a random number of photons $m$ leaving the source device, according to the Poisson distribution $p(m|\mu_k)$.
\end{enumerate}

$ $\newline
\textbf{Channel Simulation:}
\begin{enumerate}
\item For each of the $m$ photons sent by the source, choose, with probability $\gamma$ (where $\gamma$ is the expected probability of a photon surviving, according to the underwater model discussed in Section \ref{sec:uw}), whether or not that photon ``survives'' the trip from Alice to Bob (the receiver).  Let $m'$ be the total number of photons that arrive at Bob's receiver device (which may be zero, but is expected to be $m\gamma$).
\item Compute $n_N$, the number of expected background photons (noise photons) arriving at Bob's device (according to Section \ref{sec:uw}, specifically Equation \ref{eq:num-noise}). Choose random non-negative integer ``$b$'' to be the actual number of noise photons on this round (chosen using a Poisson distribution, with mean $n_N$).
\end{enumerate}

$ $\newline
\textbf{Receiver Output:}
\begin{enumerate}
  \item Let $P_C$ be the number of photons hitting Bob's ``Correct'' detector (the one that will cause him to distill a correct raw key based on Alice's key-bit choice); let $P_W$ be the number of photons hitting his ``Wrong'' detector.  From the above, along with our noise model, these are $P_C = m' + b/2$ and $P_W = b/2$.
\item The Correct detector will click with probability:
\[
Pr_C = (1-p_{dc})\times (1- (1-\eta)^{m'+b/2}) + p_{dc},
\]
where $p_{dc}$ is the probability of a dark-count in any round, and $\eta$ is the detector efficiency.  The probability the Wrong detector will click is given by:
\[
Pr_W= (1-p_{dc})\times(1- (1-\eta)^{b/2}) + p_{dc},
\]
Note that $Pr_C$ and $Pr_W$ do not add to one; indeed, both detectors may click, or none.
\item Based on the above computation, the simulator will choose a measurement outcome: Correct, Wrong, Vacuum (none clicked), or Double (both clicked).  The outcome is stored in a running total for each of the four possible outcomes and for each basis and power-level combination, allowing the simulator to construct $N_{X,i}$, $N_{Z,i}$, $M_{X,i}$, and $M_{Z,i}$, as discussed above, needed to compute Equation \ref{eq:key-size}.
\end{enumerate}


Note that we do not simulate the entire QKD stack; thus, we do not take into account the timing and operation of error correction or privacy amplification, nor do we consider communication latency involved with the classical communication required for QKD to operate.  These factors are important; however we leave that simulation as future work, and instead focus only on the final potential secret key size based on the number of communication rounds.

In our simulations, we choose power level settings that seemed to yield good results on average; further improvements may be made in optimizing over all possible settings; however this optimization is computationally expensive in our setup, due to the simulator discussed below.  For the three decoy protocol, we use $\mu_1 = .8$ (with probability $.45$), $\mu_2 = .2$ (probability $0.45$), and $\mu_3 = .004$ (chosen with probability $.1$).  For the two-decoy protocol, we use $\mu_1 = .8$ and $\mu_2 = .2$, which we choose with uniform probability.  Finally, in our simulations, we will bias towards the $Z$ basis by $20\%$.  Again, this is a setting that yielded good results in our trials; however future work may optimize over this setting.  In general, the larger $N$ is, the smaller $p_Z$ can be.

\section{Evaluation}

We evaluate the finite key performance of the decoy state protocol using our simulator discussed above, in a variety of settings.  We are particularly interested in the overall maximal supported distances between source and receiver, in various water and background light conditions.  We will consider two extreme water conditions: Pure Sea Water ($\alpha = .043$) and Coastal Ocean Water ($\alpha = .398$), the latter of which has a higher loss rate (see Table \ref{tbl:water}).  We will also consider a background light level of $E_0 = 10^{-3}W/m^2$ which represents a clear atmosphere with a full moon near the zenith,  $E_0 = 10W/m^2$ which represents heavy overcast, with the sun near the horizon, and $E_0 = 20W/m^2$ which represents a lighter overcast \cite{paglierani2023primer}.

\begin{figure}
\centering
\includegraphics[width=0.95\linewidth]{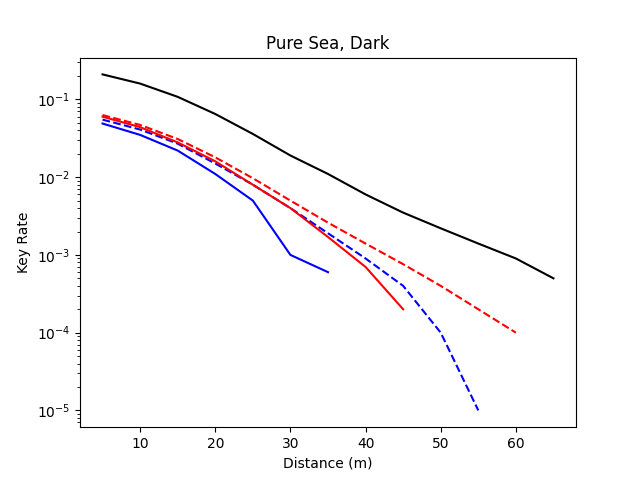}
\caption{Comparing the key-rate of the three (Blue) and the two (Red) decoy state BB84 protocol, for $N=10^7$ rounds (Solid) and $N=10^8$ rounds (Dashed)  Here we assume near ideal detector efficiency of $0.9$, pure sea water $\alpha = 0.043$, and low ambient light $E_0 = 10^{-3}$.  We note that two decoy outperforms three decoy in this scenario.  Black: Asymptotic upper bound.}\label{fig1}
\end{figure}

Figure \ref{fig1} compares the three and two decoy BB84 protocol for $N=10^7$ rounds and $N=10^8$ rounds when evaluated at night in pure sea water, using near ideal detectors, with an efficiency of $\eta = 0.9$.  We use a dark count setting of $p_d = 10^{-6}$ for all simulations.  We note that the simpler two decoy protocol always outperforms in these settings; we also note that, even at a low signal count of $N=10^7$, reasonably long distances are possible; if one is willing to perform $N=10^8$ rounds before running error correction and privacy amplification, distances approaching $60m$ are possible.  Also shown is a comparison to the expected asymptotic bound which assumes infinite time and also uses the so-called \emph{infinite decoy protocol}, where an infinite amount of decoy states are possible, thus giving Alice and Bob exact knowledge of single photon statistics.  

\begin{figure}
\centering
\includegraphics[width=0.95\linewidth]{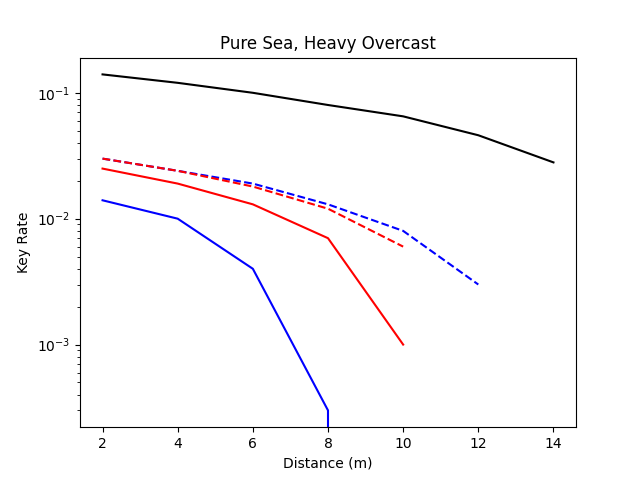}
\caption{Comparing the key-rate of the three (Blue) and the two (Red) decoy state BB84 protocol, for $N=10^7$ rounds (Solid) and $N=10^8$ rounds (Dashed)  Here we assume near ideal detector efficiency of $0.9$, pure sea water $\alpha = 0.043$, and heavy overcast, with the sun at the horizon $E_0 = 10$.  We note that at $N=10^8$ signals, three decoy outperforms slightly the two decoy protocol.}\label{fig2}
\end{figure}

\begin{figure}
\centering
\includegraphics[width=0.95\linewidth]{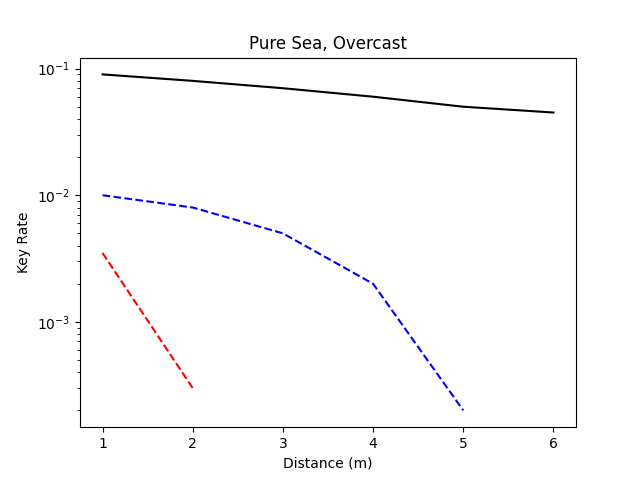}
\caption{Comparing the key-rate of the three (Blue) and the two (Red) decoy state BB84 protocol for $N=10^8$ rounds (Dashed)  Here we assume near ideal detector efficiency of $0.9$, pure sea water $\alpha = 0.043$, and lighter overcast $E_0 = 20$.  We note that, here, the three decoy state always outperforms, though distances are severely limited.  We also note that $N=10^7$ rounds did not produce a positive key-rate for either protocol.}\label{fig3}
\end{figure}

Figures \ref{fig2} and \ref{fig3} shows the same comparison but for heavy and lighter overcast respectively.   We note that in ``bright'' light, we were unable to establish a secret key with only $N=10^7$ rounds, and with $N=10^8$ rounds, we were able to establish a key, however only for relatively short distances, as shown in Figure \ref{fig3}.

\begin{figure}
\centering
\includegraphics[width=0.95\linewidth]{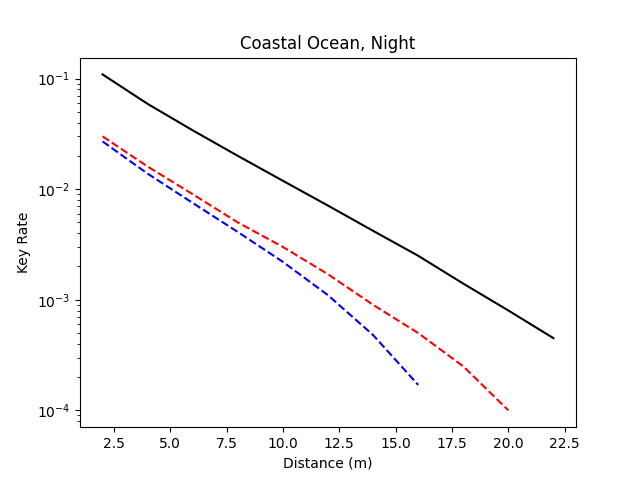}
\caption{Comparing the key-rate of the three (Blue) and the two (Red) decoy state BB84 protocol for $N=10^8$ rounds (Dashed)  Here we assume near ideal detector efficiency of $0.9$, coastal ocean water $\alpha = 0.398$, and night $E_0 =10^{-3}$.  We note that, here, again, the two decoy state protocol outperforms.  We also note that $N=10^7$ rounds did not produce a positive key-rate for either protocol.}\label{fig4}
\end{figure}

Figure \ref{fig4} compares the performance in coastal ocean water, where $\alpha = 0.398$ (i.e., a higher probability of photon loss).  We note that, here, we were unable to establish a secret key with only $N=10^7$ rounds, and were required to use $N=10^8$ rounds.  

In general, our above simulations show that, for the majority of scenarios, the two-decoy state protocol outperforms the three decoy-state protocol, especially in low noise, low loss, scenarios.  For noisier scenarios, three-decoy begins to outperform.  This is due to the fact that two decoy states are typically sufficient to get a good bound on the single photon statistics, and since a vacuum decoy is never used, more data may be transmitted \cite{rusca2018finite}.  

\begin{figure}
\centering
\includegraphics[width=0.85\linewidth]{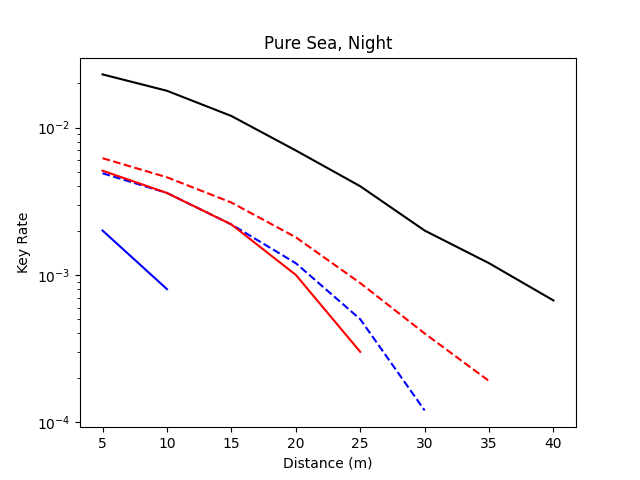}
\includegraphics[width=0.85\linewidth]{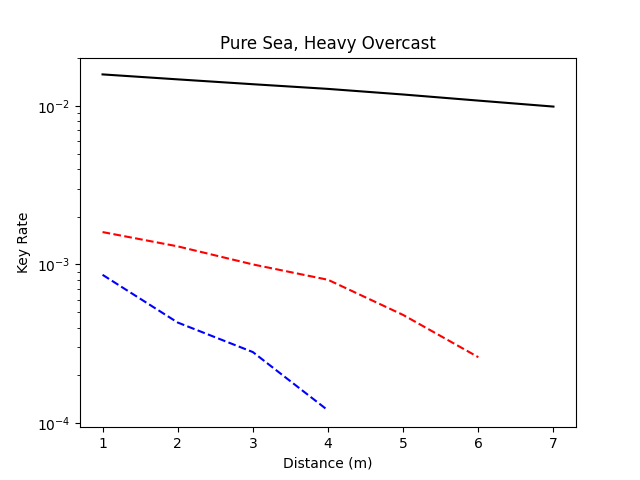}
\includegraphics[width=0.85\linewidth]{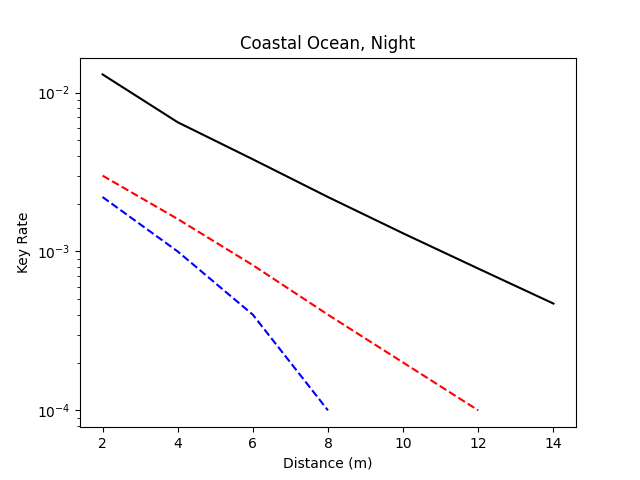}
\caption{Comparing the key-rate of the three (Blue) and the two (Red) decoy state BB84 protocol, for $N=10^7$ rounds (Solid) and $N=10^8$ rounds (Dashed)  Here we assume low detector efficiency of $0.1$. As expected, detector efficiency can greatly reduce maximal distances, even at night in ideal water.}\label{fig5}
\end{figure}

Finally, in Figure \ref{fig5}, we perform the same tests as before, but for weaker detectors, where their efficiency is only $0.1$.  As expected, maximal distances, and key-rates, drop under these scenarios, showing the importance of detector efficiency on overall underwater key rates.  Again, two decoy BB84 outperforms in general.

We note that we were unable to establish a secret key under brighter light conditions of $E_0>20W/m^2$, even in the best case of pure sea water and even if we attempt a high level of rounds, $N=10^9$.  This is due to the fact that too many background photons appear at the receiver and they ``overwhelm'' the device, creating too much noise, compared to the few number of original photons from the source end which survive the journey.  However, as we show in the next section, classical methods may help to overcome this limitation, and allow for secret key generation rates, even under such noisier scenarios.

\section{Classical Advantage Distillation (CAD)}\label{sec:cad}

We now show how CAD \cite{maurer2002secret} may be used to improve performance of the BB84 protocol underwater.  CAD is an optional process run after a raw key has been distilled, but before error correction is run.  It is meant to take a noisy raw key and produce a smaller version of it, that has fewer bit flip errors.  Thus, error correction will leak less information, which can cause the secret key length to increase (see Equation \ref{eq:key-size}).  The process, however, requires additional communication over the authenticated channel which may reveal more information to Eve.  Thus, it should only be used when the channel is very noisy --- otherwise performance degradation can be expected.

The CAD protocol we consider is one discussed in \cite{bae2007key}.  It operates on blocks of size $C \ge 2$. Alice and Bob each start with their original raw keys, produced by the QKD protocol; the goal is to output new raw keys, with fewer errors. The process begins by setting the new raw key registers for both Alice and Bob to be empty strings.  Then the following algorithm is performed until the original raw key is exhausted:
\begin{enumerate}
  \item Alice and Bob divide their raw keys into blocks of size $C$.  For every block $j$, let $a_{j,1}\cdots a_{j,C}$ be the bits of Alice's $j$'th block of $C$ bits.  Similarly, let $b_{j,1}\cdots b_{j,C}$ be Bob's raw key bits for this block.
\item For each block $j$, Alice chooses a single random bit $r_j$ (to be the candidate new raw key bit for this block).  She then sends to Bob, over the authenticated classical channel, the bit string message $a_{j,1}\oplus r_j, \cdots, a_{j,C}\oplus r_j$.  Namely, she XOR's the same bit $r_j$ with each of the $C$ bits in her block.  This communication leaks additional information to Eve potentially, which must be accounted for in the final key-rate equation.
\item Bob, on receipt of this message, which we denote $p_{j,1}\cdots p_{j,C}$, will XOR his original raw key block with this message. Namely, he will compute $b_{j,1}\oplus p_{j,1}, \cdots, b_{j,C}\oplus p_{j,C}$.  If this results in a string of all zero's or all one's, he signals to Alice to ``accept'' this block; otherwise both parties discard the block (all $C$ bits are removed from the original raw key, and no new raw key bit is added to the new raw key registers).
\item If the block is accepted, Alice adds $r_j$ to her new raw key register; Bob adds a zero or one to his new raw key register, based on his computation in the previous step.  (That is, he adds $b_{j,1} \oplus p_{j,1}$.)  Parties then discard the entire block of $C$ bits from the original raw key.
\item The above repeats for each block $j$.
\end{enumerate}

It is not difficult to see that the protocol is correct: namely, in the absence of noise, both parties will agree on the same new raw key (we will show an example momentarily).  It is also not difficult to see that the size of the new raw key will be no greater than $1/C$ of the original (it may be potentially less, since a discarded block will not add any new bit to the new raw key).  Thus, in some sense, CAD hurts performance, due to the ``compression'' of the raw key, and CAD should not always be used; in particular, when the noise in the raw key is low, CAD will substantially hurt the overall key-rate of the system.  However, if the noise is high, CAD can produce a new raw key that contains a much lower error rate, thus countering the disadvantage in raw key bit size.

The reason CAD decreases the overall noise in the new raw key is due to the fact that, now, a new raw key bit is accepted only if all $C$ bits Bob has in a block are the same or they are all different, from Alice's $C$ bits. Let's assume the average error in the original raw key is $Q$ (i.e., the probability that Alice and Bob's raw key bits do not match is $Q$).  Let's consider a block of $C$ bits.  The probability that this block is ``accepted'' (and not discarded) is $p_a = (1-Q)^C + Q^C$.  This is due to the fact that a block is accepted only if all $C$ bits are the same as Alice, or they are all different.  It is clear to see that any other case will result in Bob's XOR computation resulting in a bit-string that is not uniformly one or zero (and so will be discarded).

Now, continuing, assume the block is accepted.  If all $C$ bits of Bob are equal to Alice, then Alice and Bob will have the same, random, new raw key bit.  Otherwise, if all $C$ bits are not equal, then Alice and Bob will have a single bit error in their new raw key register.  Thus, the new expected raw key bit error rate, after CAD is run, will be $Q^C / p_a$.  It is easy to see that this is strictly smaller than $Q$ (the original raw key error rate), whenever $0 < Q < 1/2$.

To illustrate, consider the following example, where Alice's original raw key is $A = 101, 101, 010$ and $B = 101, 110, 101$.    Let's assume a CAD block size of $C=3$.

\begin{itemize}
  \item On the first block ($A_1 = 101$ and $B_1 = 101$), let's say Alice chooses $r_1 = 0$.  In this case, Alice sends to Bob the message $101$.  Bob will XOR this message with his raw key block to get $101\oplus 101 = 000$.  Since this is a string of all zeros, he signals to accept this block.  Alice's new raw key register is now $A_{new} = 0$ while Bob's is also $B_{new} = 0$.
\item On the second block, we have $A_2 = 101$ and $B_2 = 110$.  Let's say Alice chooses random $r_2 = 1$.  Thus, she sends to Bob the classical message $010$.  Bob XOR's this with his block to get $110 \oplus 010 = 100$.  Since this is not a string of all zeros or ones, he signals to discard this block.  Thus, no new raw key bit is added, and their new raw key registers remain $A_{new} = 0$ and $B_{new} = 0$ (from the first block).
\item Finally, on the third block, $A_3 = 010$ and $B_3 = 101$.  Let's say Alice chooses $r_3 = 1$ and so she sends the message $101$.  Bob XOR's this to get $101 \oplus 101 = 000$.  Since this is a string of all zeros, he signals to accept.  However, Alice will add a $1$ to her new raw key register (her choice of $r_3$), while Bob will add a zero to his.  Thus, the new raw key registers are, now, $A_{new} = 01$, $B_{new} = 00$.
\end{itemize}

The above example was constructed in order to show all three possible cases of a CAD output: Accept without error, Accept with error, or Discard.  In general, for a large raw key, the new error rate is expected to be $Q^C/(Q^C + (1-Q)^C) < Q$.

To evaluate the performance of CAD in an underwater channel, we will use key-rate equations from \cite{li2022improving} which determined a key-rate bound for the decoy state protocol with CAD.  While this was not specific to underwater channels, as with the decoy state equations discussed in Section \ref{sec:protocol}, these CAD equations are general and can be applied to any quantum channel, including an underwater one.  In the finite key setting, this is:
\begin{equation}
\text{rate} = \frac{q_{succ}N_X}{N\cdot C}\left(p_1^C H_1 - \lambda_{EC} - \Delta\right),
\end{equation}
where: $q_{succ} = Q^C + (1-Q)^C$, where $Q$ is the total error in the raw key (before CAD); $p_1 = n_1/N_X$ (using notation from Section \ref{sec:protocol}), and $H_1$ is:
\begin{align}
H_1 &= \min_{\lambda_1\cdots, \lambda_3}\left(1 - (\Lambda_0 + \Lambda_1)h\left(\frac{\Lambda_0}{\Lambda_0+\Lambda_1}\right)\right.\notag\\
&-\left. (\Lambda_2+\Lambda_3)h\left(\frac{\Lambda_2}{\Lambda_2+\Lambda_3}\right)\right)\label{eq:hae-cad}
\end{align}
is the entropy in a single photon event case (where Alice sends a single photon).  Above, we have:
\begin{align}
\Lambda_0 &= \frac{(\lambda_0+\lambda_1)^C + (\lambda_0 - \lambda_1)^C}{2p_a}\\
\Lambda_1 &= \frac{(\lambda_0+\lambda_1)^C - (\lambda_0 - \lambda_1)^C}{2p_a}\\
\Lambda_2 &= \frac{(\lambda_2+\lambda_3)^C + (\lambda_2 - \lambda_3)^C}{2p_a}\\
\Lambda_3 &= \frac{(\lambda_2+\lambda_3)^C - (\lambda_2 - \lambda_3)^C}{2p_a}
\end{align}
where $p_a = \phi_X^C + (1-\phi_X)^C$.  The minimization in Equation \ref{eq:hae-cad} is over all $\lambda_2+\lambda_3 = \phi_X$ and $\lambda_1 + \lambda_3 = \phi_X$ (here, $\phi_X$ is the single photon error rate, estimated using the decoy state method, as discussed in Section \ref{sec:protocol}).  We set $\lambda_0 = 1-\lambda_1-\lambda_2-\lambda_3$ as these values must be normalized.  Note, we assume $Z$ and $X$ basis error rates are symmetric here (which is a valid assumption in our simulator).

The new error correction leakage term, $\lambda_{EC}$ is:
\begin{equation}
\lambda_{EC} = h\left( \frac{Q^C}{Q^C + (1-Q)^C}\right),
\end{equation}
where, again, $Q$ is the error rate in the raw key distilled from the quantum communication (i.e., the raw key before CAD is run).  This is readily sampled and observed.  Finally, $\Delta$ is a finite key error term:
\begin{equation}
\Delta = 4\sqrt{\frac{C}{n}}\log_2(2\sqrt{2} + 1) \sqrt{\ln\frac{2}{\epsilon^2}} + \frac{2C}{n}\log_2\frac{1}{\epsilon},
\end{equation}
where $n$ is the size of the original raw key before CAD runs.  See \cite{li2022improving} for more details on these equations and their derivation.

\subsection{Simulation Results with CAD}

We are interested both in scenarios where, normally, standard QKD will fail to establish a raw key, however with CAD, a raw key may be established.  We are also interested in determining scenarios where CAD may actually  hurt performance.  Since the decision to turn CAD on or off, along with the exact CAD block setting, may be determined after running the quantum communication stage of the protocol (in particular, after determining the channel statistics, such as the noise and loss), it is vital to understand both scenarios.  This will provide future users of this technology guidelines on when to consider utilizing CAD, and how best to use it.  For the following, we set the number of rounds to be $N=10^8$ and test a CAD block size of two and four.

\begin{figure}
\centering
\includegraphics[width=0.95\linewidth]{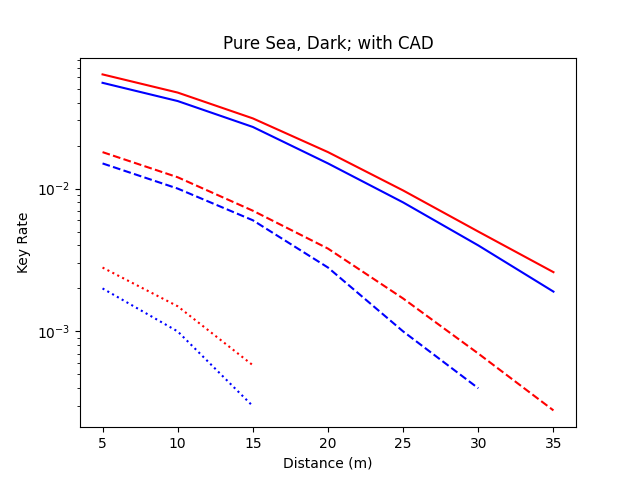}
\caption{Comparing the key-rate of the three (Blue) and the two (Red) decoy state BB84 protocol with CAD, assuming near ideal detector efficiency of $0.9$, pure sea water $\alpha = 0.043$, and a dark night $E_0 = 10^{-3}W/m^2$.  Solid lines: Results without CAD.  Dashed lines: Results with CAD block $C=2$.  Dotted lines: Results with CAD block $C=4$.  We note in this setting, CAD (non-solid lines) always performs worse than disabling CAD (solid lines).  For CAD finite key results, we use $N=10^8$.}\label{fig:cad1}
\end{figure}

\begin{figure}
\centering
\includegraphics[width=0.95\linewidth]{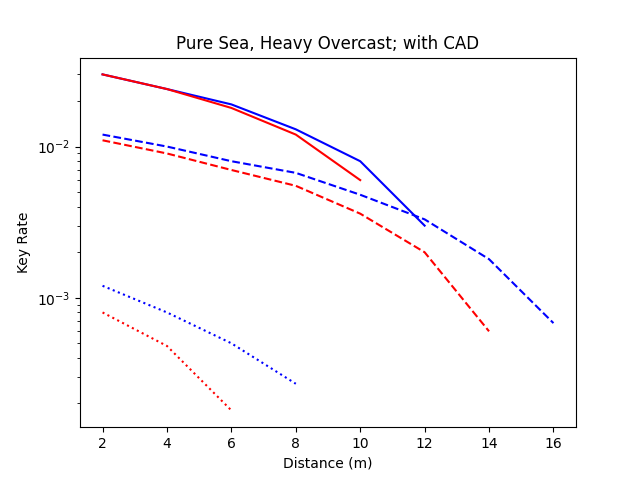}
\caption{Comparing the key-rate of the three (Blue) and the two (Red) decoy state BB84 protocol with CAD, assuming near ideal detector efficiency of $0.9$, pure sea water $\alpha = 0.043$, and a heavy overcast $E_0 = 10W/m^2$.   Solid lines: Results without CAD.  Dashed lines: Results with CAD block $C=2$.  Dotted lines: Results with CAD block $C=4$.  Here we note that setting $C=2$ allows for an increase in maximal supported distances, while $C=4$ causes worse performance.  We also note that the three-decoy protocol outperforms the two-decoy version.}\label{fig:cad2}
\end{figure}

\begin{figure}
\centering
\includegraphics[width=0.95\linewidth]{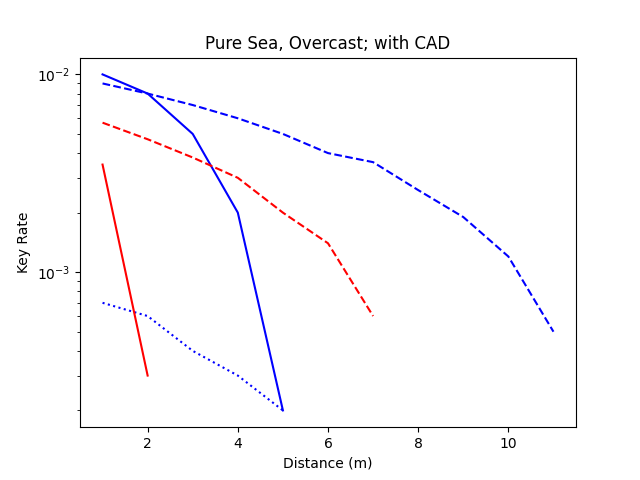}
\caption{Comparing the key-rate of the three (Blue) and the two (Red) decoy state BB84 protocol with CAD, assuming near ideal detector efficiency of $0.9$, pure sea water $\alpha = 0.043$, and a lighter overcast $E_0 = 20W/m^2$.  Solid lines: Results without CAD.  Dashed lines: Results with CAD block $C=2$.  Dotted lines: Results with CAD block $C=4$.  Here we note that setting $C=2$ allows for an increase in maximal supported distances over the case without CAD (more than doubling the maximal distance).  Again, we see $C=4$ does not help.  Note that for the Two-state protocol (Red), setting $C=4$ resulted in no secret key rate, so we did not plot that, here.}\label{fig:cad3}
\end{figure}

We first test near ideal detectors with an efficiency of $\eta = .9$.  In general, CAD will hurt performance when operating in clear water with low ambient light levels (i.e., low noise), as shown in Figure \ref{fig:cad1}.  However, when the ambient light level increases (leading to increased noise due to a higher number of background photons as discussed in Section \ref{sec:uw}), CAD can benefit performance, as shown in Figures \ref{fig:cad2} and \ref{fig:cad3}.

In more turbulent water (e.g., coastal water) we found that CAD again does not help performance in low light levels, however when the ambient light is higher, CAD can allow for a secret key to be established in up to $2m$ of water according to our simulations (we do not plot this scenario).  While this is of course a very short distance, we note that without CAD, a secret key cannot be established at $N=10^8$ rounds.  Since this distance is so short, we did not plot this data.

\begin{figure}
\centering
\includegraphics[width=0.95\linewidth]{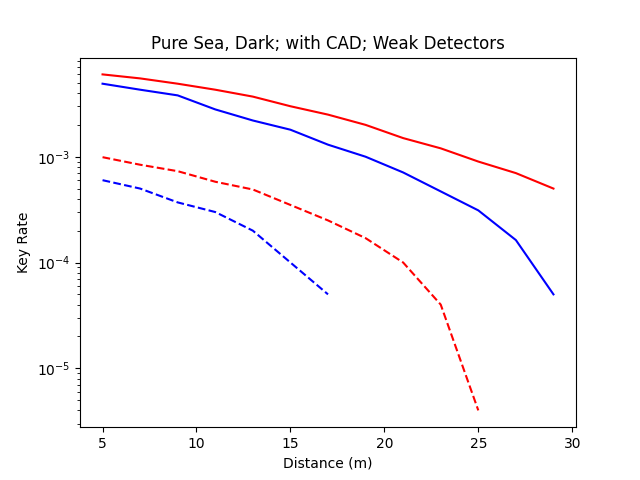}
\caption{Comparing the key-rate of the three (Blue) and the two (Red) decoy state BB84 protocol with CAD, assuming weak detectors, with an efficiency of $0.1$, pure sea water $\alpha = 0.043$, and low ambient light $E_0 = 10^{-3}W/m^2$.  Solid lines: Results without CAD.  Dashed lines: Results with CAD block $C=2$.  We see that CAD does not improve performance under these conditions.}\label{fig:cad4}
\end{figure}

For weak detectors with an efficiency of $\eta = .1$, we found that CAD never helped, when the signal count was $N = 10^8$.  Figure \ref{fig:cad4} shows the case of low ambient light.  For other tests, CAD did not produce a positive key-rate, and so we do not plot them.  Positive key-rates were possible with CAD, when $N=10^{-9}$, however our simulation was too slow to plot these.  Instead, we will comment on the asymptotic results, below.

Note that in our simulations, setting $C=4$ never helped performance in our evaluations.  We found this to be  due to the fact that our round number was too low ($N=10^8$).  A higher number of rounds slowed the simulation substantially and, furthermore, $N=10^8$ is already a high setting for QKD in the finite key setting \cite{lim2014concise}.  However, since CAD has not been evaluated at all in the underwater scenario, we also consider the asymptotic performance of the system.

Computing asymptotic results (where $N\rightarrow \infty$) and using expected error rates and photon loss as determined by the underwater model, described in Section \ref{sec:uw}), we see in Table~\ref{tbl:cad-asym} that higher CAD block sizes can benefit performance in higher light levels when the loss is low (i.e., over Pure Sea Water).  When the loss is high (see Table \ref{tbl:cad-asym-2}), CAD can still help; however $C=2$ seems to be the optimal choice, again.  We are also able to test this in turbulent harbor water (the water setting, in Table \ref{tbl:water}, with the highest attenuation coefficient).  In the finite key setting, we were unable to establish a secret key under these conditions; asymptotically, we see it is possible, though distances are severely limited; CAD, however, does help slightly.  These results are shown in Table \ref{tbl:cad-asym-3}.  In this table, we evaluate both  near ideal detectors (with an efficiency of $90\%$) and less ideal detectors (with a $10\%$ efficiency).

In general, according to our simulated results, in practice setting $C=2$ provides the greatest benefit to overall performance in noisy scenarios.  We also see that CAD does not help in more turbulent water (which makes sense, since CAD is specifically designed to assist with noisy channels, not lossy ones).  We also see that CAD is sensitive to the number of signals sent.  If one is able to perform a higher number of QKD rounds, then setting CAD to a higher block setting may be helpful if the loss is not too high.

\begin{table}
\caption{Showing the maximal asymptotic distance for various light levels and CAD block settings, when operating in Pure Sea Water ($\alpha = 0.043$).  Interestingly, with weaker detectors, slightly higher distances are possible in brighter light, since the background photons are not all detected due to the detector imperfections, though the increase is slight.}\label{tbl:cad-asym}
\centering
\begin{tabular}{l|cccc}
\multicolumn{5}{c}{Ideal Detectors ($\eta = .9$)}\\
\hline
Light Level ($E_0$) & No CAD & $C=2$ & $C=4$ & $C=5$\\
\hline
$0.001$ W/$m^2$ & 125m & 135m & 115m & 105m\\
$10$ W/$m^2$ & 17m & 19m & 21m & 21m\\
$20$ W/$m^2$ & 11m & 13m & 15m & 16m\\
\hline\hline
\multicolumn{5}{c}{Weak Detectors ($\eta = .1$)}\\
\hline
$0.001$ W/$m^2$ & 115m & 115m & 90m & 80m\\
$10$ W/$m^2$ & 18m & 22m & 24m& 25m\\
$20$ W/$m^2$ & 10m & 15m & 18m & 19m
\end{tabular}
\end{table}

\begin{table}
\caption{Showing the maximal asymptotic distance for various light levels and CAD block settings, when operating in Coastal Ocean Water ($\alpha = 0.398$).}\label{tbl:cad-asym-2}
\centering
\begin{tabular}{l|cccc}
\multicolumn{5}{c}{Ideal Detectors ($\eta = .9$)}\\
\hline
Light Level ($E_0$) & No CAD & $C=2$ & $C=4$ & $C=5$\\
\hline
$0.001$ W/$m^2$ & 33m & 34m & 32m & 30m\\
$1$ W/$m^2$ & 11m & 13m & 12m & 12m\\
$5$ W/$m^2$ & 5m & 7m & 6m & 6m\\
\hline\hline
\multicolumn{5}{c}{Weak Detectors ($\eta = .1$)}\\
\hline
$0.001$ W/$m^2$ & 32m & 31m & 27m & 24m\\
$1$ W/$m^2$ & 11m & 13m &13m & 13m\\
$5$ W/$m^2$ & 5m & 7m &8m & 8m
\end{tabular}
\end{table}

\begin{table}
\caption{Showing the maximal asymptotic distance for various light levels and CAD block settings, when operating in Turbid Harbor Water ($\alpha = 2.190$).}\label{tbl:cad-asym-3}
\centering
\begin{tabular}{l|cccc}
\multicolumn{5}{c}{Ideal Detectors ($\eta = .9$)}\\
\hline
Light Level ($E_0$) & No CAD & $C=2$ & $C=4$ & $C=5$\\
\hline
$0.001$ W/$m^2$ & 4.8m & 5m & 4.4m & 4.2m\\
$1$ W/$m^2$ & $1.2$m & $1.4$m & $1.6$m & $1.6$m\\
$5$ W/$m^2$ & $0.4$m & $0.6$m & $0.8m$ & $0.8$m\\
\hline\hline
\multicolumn{5}{c}{Weak Detectors ($\eta = .1$)}\\
\hline
$0.001$ W/$m^2$ & $4.6$m & $4.4$m & $3.6$m & $3.2$m\\
$1$ W/$m^2$ & $1.2$m & $1.4$m & $1.6$m & $1.4$m\\
$5$ W/$m^2$ & $0.4$m & $0.6$m & $0.8$m & $0.8$m
\end{tabular}
\end{table}

\section{Closing Remarks}

In this paper, we evaluated, for the first time, the expected performance of underwater QKD in the realistic finite key scenario.  Past work only considered asymptotic bounds.  We also tested two different decoy state variants and showed when it was optimal to use the ``simpler'' protocol over the standard three-decoy version.  Finally, we also investigated the use of classical advantage distillation, to improve the maximal distances over which QKD can establish a secret key underwater.  Prior to our work, the use and behavior of classical post-processing methods were not investigated in underwater channels.  Our work shows that it is vital to investigate alternative protocols, and also alternative classical post-processing strategies, to allow for secret key generation over a wide variety of operating conditions underwater.

Many interesting future problems remain.  While we showed that CAD can improve performance in daylight conditions, it did not help in high loss conditions.  Investigating alternative strategies (either alternative QKD protocols, or alternative classical strategies) would be beneficial so as to allow QKD to operate in more turbulent water.  Designing a simulator capable of running beyond $N=10^8$ rounds, efficiently, would also be beneficial, especially when evaluating CAD results.  Finally, we did not take into account latency in the classical network (needed for sampling, sifting, and CAD communication).  It would be interesting to design a more complete simulation environment, to take these factors into account.

\section*{Acknowledgments}
TT, SP, and WOK would like to acknowledge support from the Department of Navy award N00014-24-1-2093 issued by the Office of Naval Research.  SP performed this work while at the University of Connecticut.  Any opinions, findings, and conclusions or recommendations expressed in this material are those of the author(s) and do not necessarily reflect the views of the Office of Naval Research.

\bibliographystyle{unsrt}
\bibliography{new-bib}
\end{document}